\documentclass[twocolumn]{aastex631}
\usepackage{natbib}
\usepackage{appendix}
\usepackage{multirow}
\usepackage{longtable}
\begin{document}

\title{Statistical Distribution Function of Orbital Spacings in Planetary Systems}

\correspondingauthor{Jeremy Dietrich}
\email{jdietrich@asu.edu}

\author[0000-0001-6320-7410]{Jeremy Dietrich}
\affiliation{Department of Astronomy, The University of Arizona, Tucson, AZ 85721, USA}
\affiliation{School of Earth and Space Exploration, Arizona State University, Tempe, AZ 85287, USA}

\author[0000-0002-1226-3305]{Renu Malhotra}
\affiliation{Lunar and Planetary Laboratory, The University of Arizona, Tucson, AZ 85721, USA}

\author[0000-0003-3714-5855]{D\'aniel Apai}
\affiliation{Department of Astronomy, The University of Arizona, Tucson, AZ 85721, USA}
\affiliation{Lunar and Planetary Laboratory, The University of Arizona, Tucson, AZ 85721, USA}

\begin{abstract}
The minimum orbital separation of planets in long-stable planetary systems is often modeled as a step function, parameterized with a single value $\Delta_{min}$ (measured in mutual Hill radius of the two neighboring planets). Systems with smaller separations are considered unstable, and planet pairs with greater separations are considered stable. Here we report that a log-normal distribution function for $\Delta_{min}$, rather than a single threshold value, provides a more accurate model. From our suite of simulated planetary systems, the parameters of the best-fit log-normal distribution are $\mu=1.97\pm0.02$ and $\sigma=0.40\pm0.02$, such that the mean, median, and mode of $\Delta_{min}$ are 7.77, 7.17, and 6.11, respectively. This result is consistent with previous estimates for $\Delta_{min}$ threshold values in the range $\sim$5$-$8. We find a modest dependence of the distribution of $\Delta_{min}$ on multiplicity within the system, as well as on planetary mass ratios of the closest planet pair. The overall distribution of nearest-neighbor planetary orbital spacings (measured in the mutual Hill radii and denoted simply as $\Delta$) in long-term stable systems is also well fit with a log-normal distribution, with parameters $\mu=3.14\pm0.03$ and $\sigma=0.76\pm0.02$. In simulations of sets of many planets initially packed very close together, we find that the orbital spacings of long-term stable systems is statistically similar to that in the observed Kepler sample of exo-planetary systems, indicating a strong role of sculpting of planetary architectures by dynamical instabilities.
\end{abstract}



\section{Introduction} \label{sec:intro}

NASA's \textit{Kepler} mission \citep[][]{Borucki2010} has discovered a very large number of exoplanetary systems that can be studied in a consistent manner for planetary system demographics. One of the properties of particular interest for planetary system demographics is the orbital spacings (equivalently, orbit period ratios) of nearest-neighbor planets. Studies of the \textit{Kepler} population statistics have shown that the distribution of orbital period ratios of adjacent planets has an overall broad peak and several narrow peaks associated with low-integer orbital period ratios \citep[][]{Fabrycky2014,Malhotra2015,Mulders2018}. The distribution of orbital spacings observed in the \textit{Kepler} sample peaks near 10--20 mutual Hill radii, with significant dependence on multiplicity \citep[][]{Pu2015,Gilbert2020}. There is also a long tail of wide spacings, indicative of large gaps between planets where additional undetected planets may exist \citep[e.g.,][]{Gilbert2020, Dietrich2020}. 
\citet[][]{He2019} generated sets of simulated ``Kepler-like" planetary systems, simulated observing them with the Kepler detection pipeline, and reported that the best-fit models to the actual observational data requires minimum orbital spacings of about 8 mutual Hill radii.
In the present study, our goal is to obtain a better characterization of the statistical distribution of the minimum orbital spacing of planets in long term stable systems, including its dependence on the mass ratios of the planets and on planet multiplicity in the system.

We will use the parameter $\Delta$ defined as the orbital spacing in units of the mutual Hill radius:
\begin{equation}
    \Delta = \left(\frac{2(a_2 (1 - e_2) - a_1 (1 + e_1))}{a_1 + a_2}\right)\left(\frac{3M_*}{m_1 + m_2}\right)^{1/3},
    \label{e:Delta}
\end{equation}
where $a$ is the semi-major axis, $e$ is the eccentricity, $i$ is the mutual inclination, and $m$ is the mass of each planet. The subscripts $1$ and $2$ indicate the inner and outer planet, respectively, and $M_*$ is the mass of the host star. (Note that the inclination does not appear in this definition.) Other studies also denote this dynamical spacing parameter as $K$ \citep[see e.g.,][]{Pu2015, Malhotra2015}. We will specifically use the notation ``$\Delta_{min}$" for the minimum value of $\Delta$ taken over all nearest-neighbor planet pairs in a planetary system; unless otherwise indicated, this will be the value at an initial epoch. For a system of two planets of initially circular and coplanar orbits, the theoretical minimum value of $\Delta$ for Hill stability is given by \citep[][]{Birn1973,Gladman1993,Tremaine2023}
\begin{equation}
    \Delta^*_{min} = 2\sqrt{3} \approx 3.46
    \label{e:Deltaminstar}
\end{equation} 
In the more general case, there is no known simple, analytic stability criterion for multi-planet systems, therefore insights have been sought with numerical simulations.

Numerical studies of the dynamical stability of multi-planet systems do not seek a definitive statement over all time, but only over a finite time range, usually limited by computational resources. In modern times, such investigations have extended to timescales of $10^7-10^9$ orbital periods of the innermost planet in the system \citep[e.g.,][]{Chambers1996,Smith2009}. Numerical techniques that reduce computational costs and reliably predict long-term stability based on the short-term behaviors of orbital parameters (inclination, eccentricity, and angular momentum) are also being developed \citep[see e.g.,][]{Tamayo2020, Volk2020}.

In reviewing the literature on numerical simulations of the stability of multi-planet systems, \citet[][]{Malhotra2015} reported that for systems of 3--20 equal-mass planets in initially coplanar and nearly circular orbits, stability times exceeding $10^8$ orbital periods require $\Delta_{min}\geq 8$  for low mass planets (planet-to-star mass ratio in the range $10^{-9}$--$10^{-5}$), whereas systems of higher planet masses (planet-to-star mass ratio in the range $10^{-3.4}$--$10^{-2.4}$) are stable with somewhat smaller minimum separations, $\Delta_{min}\geq 5$.

In almost all previous modeling studies, the minimum value of $\Delta$ required for long-term stability has been treated as a simple cutoff, where systems with $\Delta_{min}$ below some threshold are considered unstable and systems with $\Delta_{min}$ above that threshold considered stable. However, even a cursory examination of previous numerical results shows that a single cutoff value is merely an approximation, and that this threshold is not a single valued parameter but has a range of values that may depend upon many characteristics of a planetary system, such as the stellar mass, the number of planets, their masses, orbital eccentricities, mutual inclinations, their ordering, and the mass ratio of the planet pair with the minimum orbital separation. Because there are potentially many dependencies on parameters that are themselves random variates, we make the reasonable hypothesis that the threshold value of $\Delta_{min}$ can be better characterized as a random variate whose distribution can be illuminated with N-body numerical simulations. 

Our approach is to carry out N-body numerical simulations of a large set of simulated planetary systems with close-packed planets, measure the minimum dynamical separation, $\Delta_{min}$, in the long-term stable systems, and thereby quantitatively determine a more accurate characterization of the distribution of the threshold value of this parameter. This paper is organized as follows. We describe the methods we used to create the synthetic planetary system populations in Section~\ref{sec:methods}. The results of the N-body integrations to assess the dynamical stability of these systems are described in Section~\ref{sec:results}. In Section~\ref{sec:discussion} we discuss our results for the distribution of $\Delta_{min}$ and its dependence on planet mass and multiplicity. We summarize our findings in Section~\ref{sec:summary}.

\section{Methods} \label{sec:methods}

Our general process for determining the distribution of the minimum stable value $\Delta_{min}$ with simulated planetary systems is the following:
\begin{enumerate}
    \item We generate our initial planetary systems based on prior model distributions for planet multiplicity, orbital period, and planet mass.
    \item We use N-body integrations to evolve each system in time for 1 million orbits of the innermost planet and measure the parameters at evenly spaced time intervals.
    \item Systems that become unstable (either via ejection, collision, or final orbits that lead to instability at later times) are separated.
    \item The systems (both stable and unstable) are ordered based on the minimum initial $\Delta$ value $\Delta_{min}$ from the original system.
    \item We also search for correlations between $\Delta_{min}$ and the period and mass ratios of the corresponding planet pair, as well as the planet multiplicity of the system.
\end{enumerate}

To generate our simulated planetary systems, we needed to initialize the number of planets in a system as well as the orbital periods, masses, inclination, and eccentricity for each of the planets. The distribution of orbital periods between planet pairs is motivated by the Kepler population of exoplanets \citep[][]{Mulders2018}. The Kepler sample from \citet[][]{Mulders2018} includes planets with orbital periods in the range [2, 400] days and planet radii in the range [0.5, 6] $R_\oplus$, with cuts on the stellar population to remove giant and sub-giant stars via the effective temperature dependent surface gravity criterion \citep[][]{Huber2016}. This led to 1201 total planets in 487 multi-planet systems, with an additional 1840 observed single systems. The other parameters were initialized in a wide range so as to account for as many different planetary system architectures as possible. Specifically, we chose the following parameters and ranges for each simulated system:
\begin{itemize}
    \item \textbf{Number of planets}: randomly chosen with a uniform prior distribution between 3 and 10 planets in the system.
    \item \textbf{Orbital periods}: the first planet's orbital period randomly chosen with a uniform prior in log period space between 1 to 100 days. Additional planets had their period ratio with the previous planet drawn from the distribution of observed pairwise period ratios in the Kepler multi-planet systems, which are taken from the sample in \citet[][]{Mulders2018}. 
    \item \textbf{Masses}: randomly chosen for each planet with a uniform prior in log mass between 1/3 Earth mass and 3 Jupiter masses.
    \item \textbf{Inclination and Eccentricity}: We set each planet to have a coplanar and circular orbit, but these parameters are allowed to evolve due to interactions during the N-body integrations.
    \item \textbf{Angular Variables}: The longitude of ascending node, argument of periastron, and mean anomaly were initialized for each planet randomly from a uniform distribution across the range [0, 2$\pi$].
\end{itemize}

Based on provisional results for the initial distributions of $\Delta_{min}$, we generated two different sets of simulated planetary systems. The most significant difference between the two sets is how closely packed the systems are at the initial time. The first set, identified as the \textit{Nominal Kepler Analog Set}, uses the period ratio distribution from Kepler as described above, whereas the second set -- the \textit{Very Closely Packed Set} -- increases the number of neighboring planet pairs with $\Delta \lesssim 8$.

\subsection{Nominal Kepler Analog Set} \label{subsec:init}

\begin{figure}[t]
    \centering
    \includegraphics[width=\columnwidth]{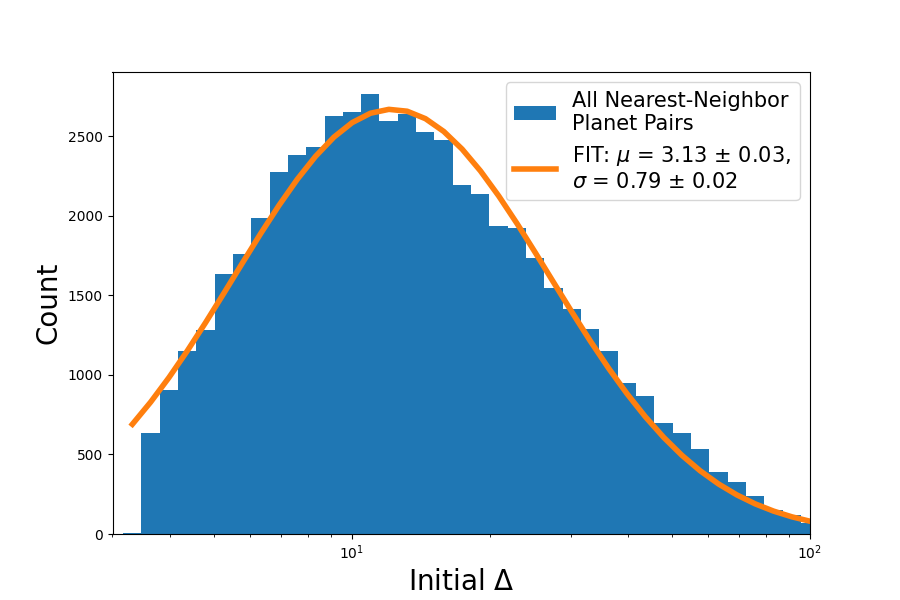}
    \caption{The distribution of initial $\Delta$ (nearest-neighbor orbital separations in units of mutual Hill radius) of the Nominal Kepler Analog Set of simulated planetary systems. The random draws we performed generated a roughly log-normal distribution of $\Delta$ in a system, as shown by the overplotted fit.}
    \label{fig:all_deltas}
\end{figure}

We generated a set of 10,000 simulated planetary systems using the procedure described above and calculated $\Delta$ between each of the nearest-neighbor planet pairs (see Figure~\ref{fig:all_deltas}). This set of draws for the planetary parameters yields an approximately log-normal distribution of $\Delta$ with parameters $\mu=3.13\pm0.03$ and $\sigma=0.79\pm0.02$. The log-normal distribution has the following probability density function,
\begin{equation}
    f(x) = \frac{1}{x\sigma\sqrt{2\pi}}\exp\left(-\frac{(\ln x - \mu)^2}{2\sigma^2}\right),
\end{equation}
such that the natural logarithm of $x$ is normally distributed with mean $\mu$ and standard deviation $\sigma$. Note that we only use ``log" as the natural logarithm within the name of this distribution; for all other uses ``log" refers to the base-10 logarithm. This functional form was previously introduced by \citet[][]{Malhotra2015} in an empirical way for the distribution of dynamical spacings of planetary systems. We also tested the beta, exponential, gamma, Lorentzian, and skewed normal distributions, but found that the log-normal function was the best fit as measured by residual sum of squares and/or reduced chi-square metrics. The best fit model parameters and their uncertainties are derived using the Python module \texttt{lmfit} to fit the distribution, with up to 600 repetitions ($200 \times n$ for $n = 3$ parameters in the fit: amplitude, $\mu$, and $\sigma$) using the Levenberg-Marquardt algorithm \citep[][]{Levenberg1944, Marquardt1963}. The modest but visible mismatch between the fit and the distribution is a result of the hard lower limit we placed on the initial $\Delta$ values (see Eq.~\ref{e:Deltaminstar}).  If we remove that limitation, a small fraction ($\sim$2\%) of additional systems are generated with $\Delta < \Delta_{min}$, mostly removing the mismatch.

\subsection{Very Closely Packed Set} \label{subsec:packed}

Our initial setup (see previous paragraph) yielded a low number of closely separated nearest neighbor planet pairs (i.e., planet pairs with $\Delta < 8$) due to the distribution of Kepler period ratios peaking at $\sim$1.9. This is an understandable result of a survival bias: The Kepler period distribution is derived from the observational data of systems that are presumably long-term stable if we observe them now, and thus few are marginally unstable. Therefore, to mitigate the bias against low values of $\Delta$ in our Nominal Kepler Analog Set, we also generated additional systems to fill in the low $\Delta$ bins to obtain a nearly uniform distribution in $\log\Delta$ at $\Delta \lesssim 10$; the latter is the value of $\Delta_{min, Kepler}$ \citep[][]{Mulders2018}. While this overly tuned our simulated planetary systems to be very closely packed, it allowed us to determine if the $\Delta_{min}$ distribution of the dynamically stable systems that survive until the end of our N-body integrations are determined more by dynamical stability itself or by the bias in the prior distribution of planets. Our procedure for generating the additional systems was as follows:

\begin{itemize}
    \item We generated 10,000 additional systems based on the same criteria as above, except the period ratios were drawn from a uniform distribution in the range [1, 3], with the imposed condition that all resulting $\Delta$ values, for every nearest-neighbor planet pair in each system, were less than 10.
    \item We accepted a system only if it did not push the number of systems for the $\Delta$ bins that were much lower than the peak of the initial distribution (at $\Delta \approx 10$) above that peak value in the histogram. This filled in much of the left side of the initial distribution, but still produced very uneven results in some of the bins in log $\Delta$ due to low number statistics from the low probability to generate certain planet pair parameter combinations.
    \item To achieve a more even range across all the bins, we again generated additional systems with the same planet multiplicity and planet mass criteria, except that we further limited the period ratio draws to only produce $\Delta$ values within the bins with fewer systems.
\end{itemize}

\begin{figure}[t]
    \centering
    \includegraphics[width=\columnwidth]{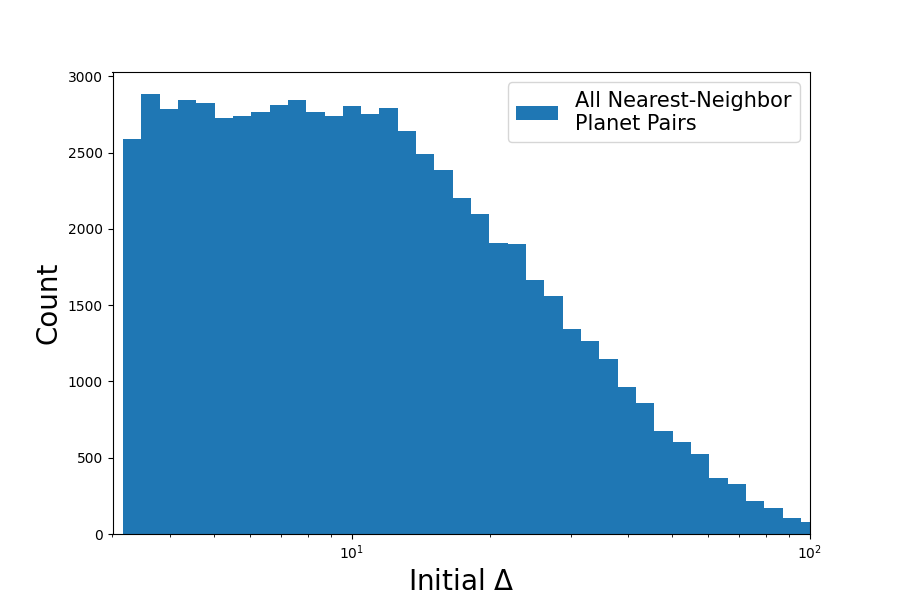}
    \caption{The distribution of initial $\Delta$ (nearest-neighbor orbital separations in units of mutual Hill radius) of the Very Closely Packed Set of simulated planetary systems. Note that this distribution is roughly uniform in the range $\sim$3.5 -- 10 (compare with Figure~\ref{fig:all_deltas})}.
    \label{fig:low_deltas}
\end{figure}

This method of generating simulated planetary systems yielded a relatively uniform initial distribution (see Figure~\ref{fig:low_deltas}) of $\log \Delta_{min}$ in the range $\Delta = 2 \sqrt{3}$ (the theoretical minimum separation for Hill stability of two planets) to $\Delta \sim 10$ (the peak of $\Delta_{min}$ of the Nominal Kepler Analog Set). Of the other variables we track (multiplicity, period ratios, mass ratios), this procedure mostly affected only the period ratio distribution by increasing the number of low period ratios between the nearest-neighbor planet pairs. The multiplicity distribution and mass ratio distribution between nearest-neighbor planets remained similar to those from the Nominal Kepler Analog set.

\subsection{N-body Integrations} \label{subsec:Nbody}

We ran each system for both sets of simulated planetary systems through N-body integrations using the \texttt{mercurius} integrator in the \texttt{REBOUND} package \citep[][]{Rein2012, Rein2019}. We integrated each system over 5 million orbits of the innermost planet, recording the orbital elements at 3,000 evenly spaced time intervals along the way. At the end of those integrations, we identified each system as unstable or stable based on whether collisions (two planets' physical radii overlapping) and/or ejections occurred or did not occur during the time span of the integration. We note that stability for up to 5 million orbits is not necessarily indicative of long-term stability on the timescale of billions of years, but this timescale of integration allows to run thousands of N-body simulations over a reasonable computational time period.

For the systems from both sets that were deemed stable (i.e., experiencing no collisions or ejections), we also ran the spectral fraction analysis designed by \citet[][]{Volk2020} to predict longer-term stability from short-term numerical integrations. This analysis computes the power spectrum of the orbital parameters in the exoplanet system propagated for timescales of $\sim 10^6$ orbits of the innermost planet and looks for empirical indicators of long-term stability in the power spectrum, specifically the fraction of the spectral frequencies having power above a certain threshold fraction of the power of the peak frequency. The best empirical measure of stability was found to be as follows: with a threshold value of 5\% of the peak power, a spectral fraction below 1\% is indicative of long-term stability (on at least $(10^9)$ orbital periods), whereas a larger spectral fraction is very likely long-term unstable.

We also checked the stability criterion as measured by SPOCK \citep[][]{Tamayo2020}, which uses shorter N-body integrations ($sim 10^4$ orbits) and machine learning algorithms trained on typical stable systems to predict long-term stability. In general, we found that the SPOCK analysis found more systems unstable than the spectral fraction analysis or the $\sim 10^6$ orbit N-body survival rate, similar to results on a single system found by \citet[][]{Dietrich2022}. Unsurprisingly, SPOCK found a much higher fraction of the Very Closely Packed sample to be unstable compared with the Nominal Kepler Analog sample; this is because the former includes a much larger population of low $\Delta$ systems.

\section{Results} \label{sec:results}

For both the Nominal Kepler Analog Set and the Very Closely Packed Set, the overall fraction of unstable systems is less than 50\%, but significantly larger in the latter set. The Nominal Kepler Analog Set had an unstable system fraction of $\sim$11\%, whereas the Very Closely Packed Set had an unstable system fraction of $\sim$37\%. The unstable cases for both sets of simulated planetary systems are concentrated around a $\Delta_{min}$ $\lesssim$5 within the system. The spectral fraction analysis found a small subset ($\leq$5\%) of originally stable systems to be classified as unstable in both the Nominal Kepler Analog Set and the Very Closely Packed Set.

\subsection{Nominal Kepler Analog Set} \label{subsec:res_init}

\begin{figure*}[ht]
    \centering
    {\Large \textbf{Stability in the Nominal Kepler Analog Set}}\\
    \includegraphics[width=\columnwidth]{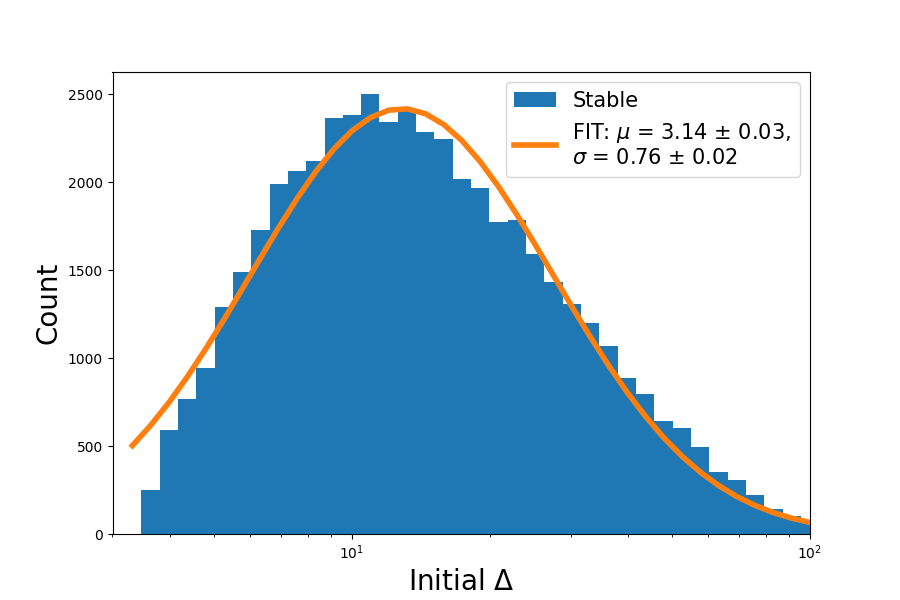}
    \includegraphics[width=\columnwidth]{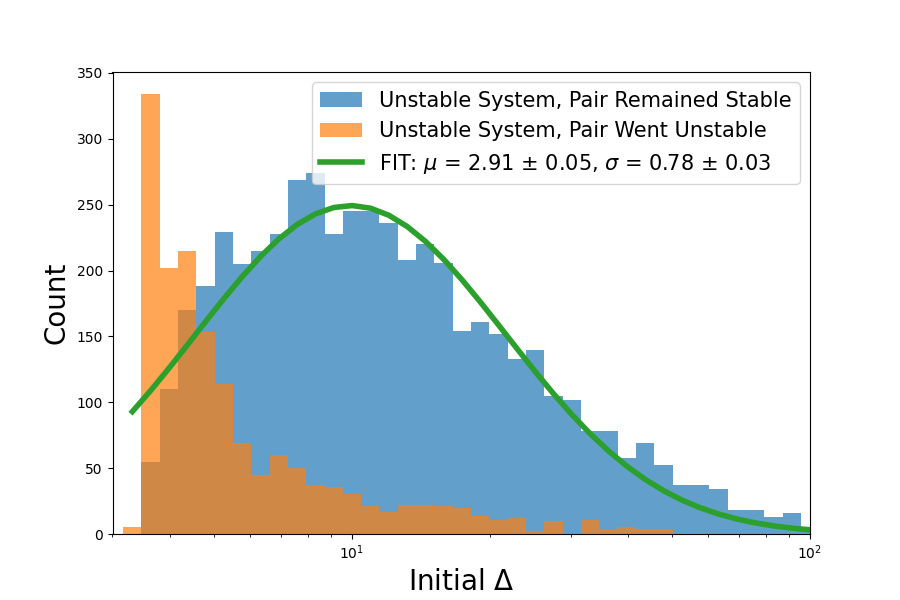}
    \includegraphics[width=\columnwidth]{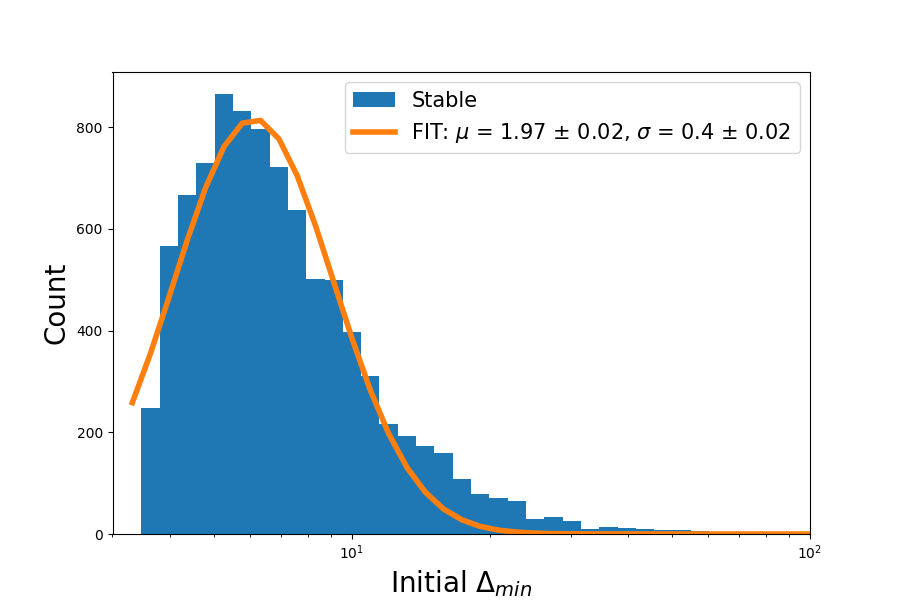}
    \includegraphics[width=\columnwidth]{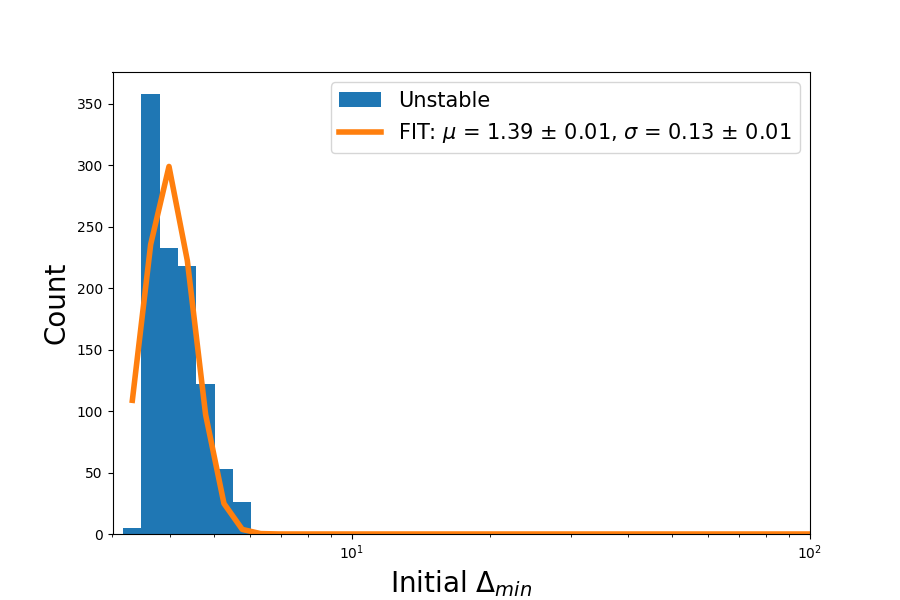}
    \caption{Results for the Nominal Kepler Analog Set. Top: Histogram of all the $\Delta$ values for each nearest-neighbor planet pair in stable systems (left) and unstable systems (right). Bottom: Histogram of the initial $\Delta_{min}$ distribution and the best-fit log-normal function for stable systems (left) and unstable systems (right).}
    \label{fig:orig1}
\end{figure*}

We calculated the best-fit initial distributions of $\Delta$ and $\Delta_{min}$ for both the stable and unstable planetary systems, as seen in Figure~\ref{fig:orig1}. The stable systems' initial $\Delta$ distribution essentially matched the distribution of all systems in this set, as over 90\% of systems were found stable. The $\Delta_{min}$ distribution most closely matched a log-normal distribution with parameters $\mu=1.97\pm0.02$ and $\sigma=0.40\pm0.02$. These values correspond to a median value of $\Delta_{min}$ of $7.17\pm0.15$, a root-mean-square value $\langle\Delta_{min}^2\rangle^{\frac{1}{2}} = 8.4\pm0.3$, and the range $(4.8, 10.7)$ contains two-thirds of $\Delta_{min}$ values. The distribution for the unstable systems' initial $\Delta_{min}$ was also fit to a log-normal distribution, with parameters $\mu=1.39$ and $\sigma=0.13$
.

\begin{figure*}[ht]
    \centering
    {\Large \textbf{Stability in the Nominal Kepler Analog Set}}\\
    \includegraphics[width=\columnwidth]{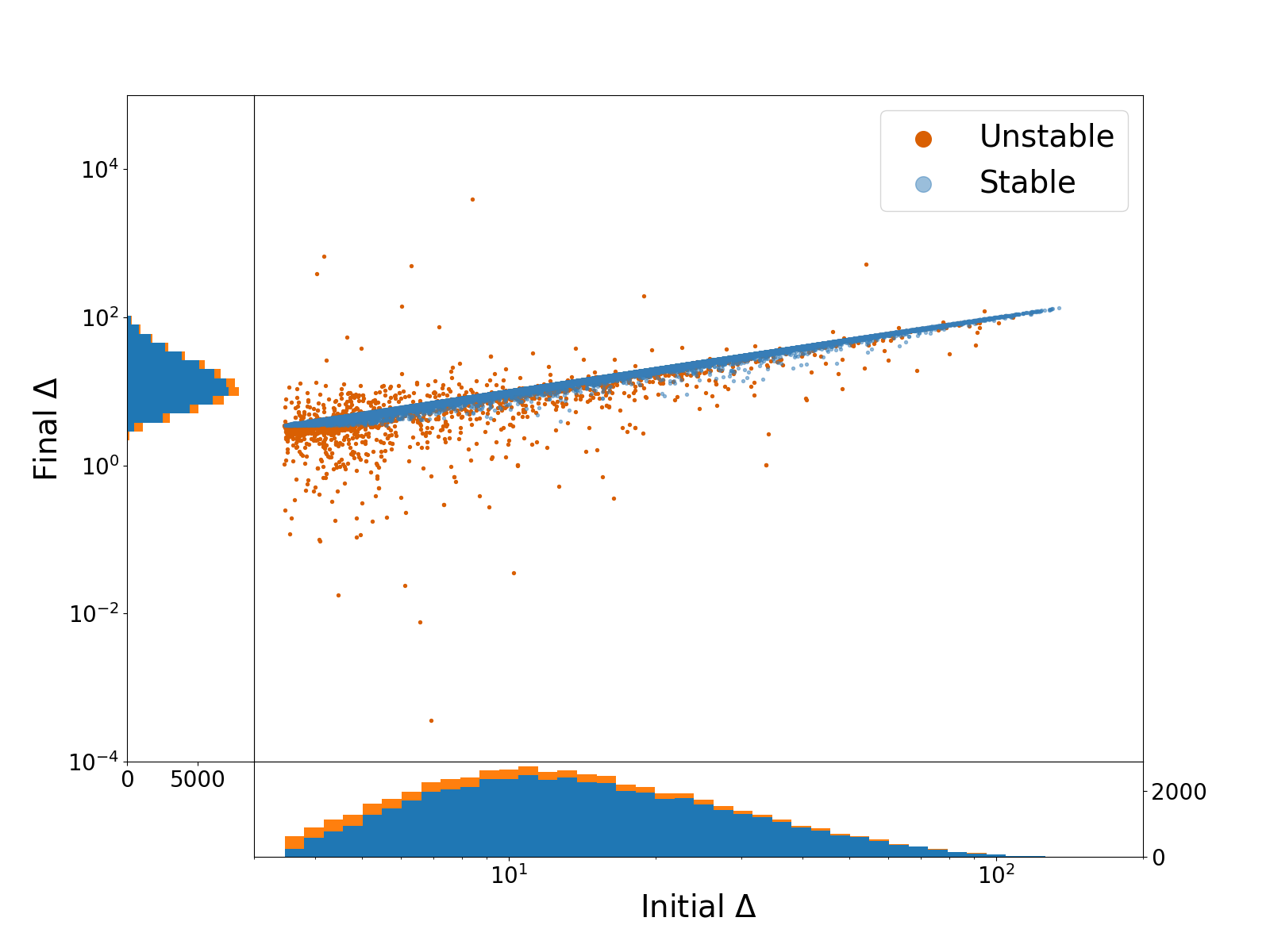}
    \includegraphics[width=\columnwidth]{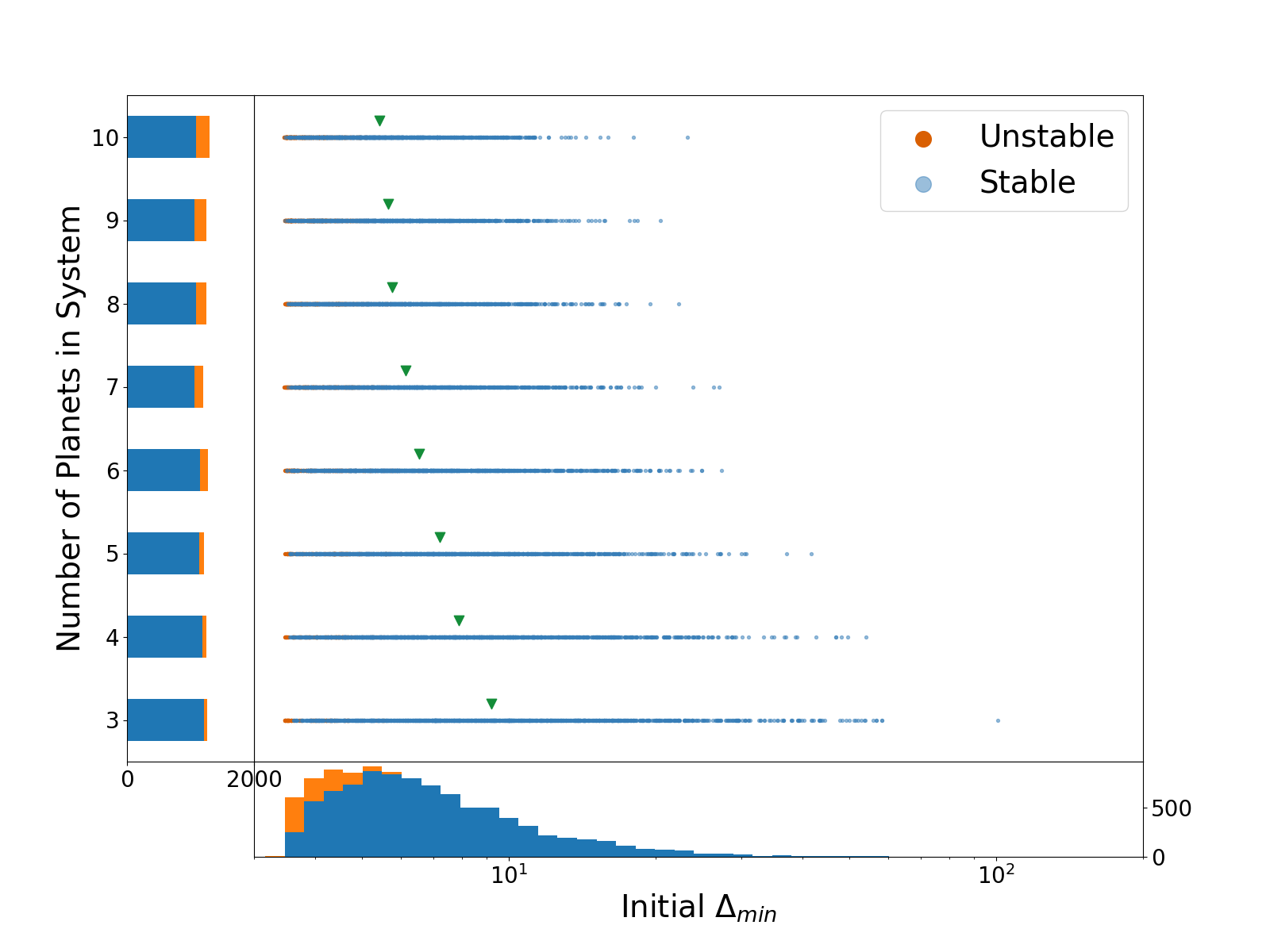}
    \includegraphics[width=\columnwidth]{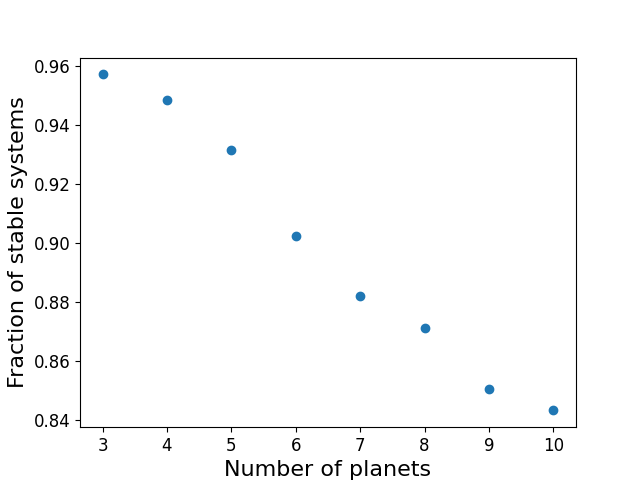}
    \includegraphics[width=\columnwidth]{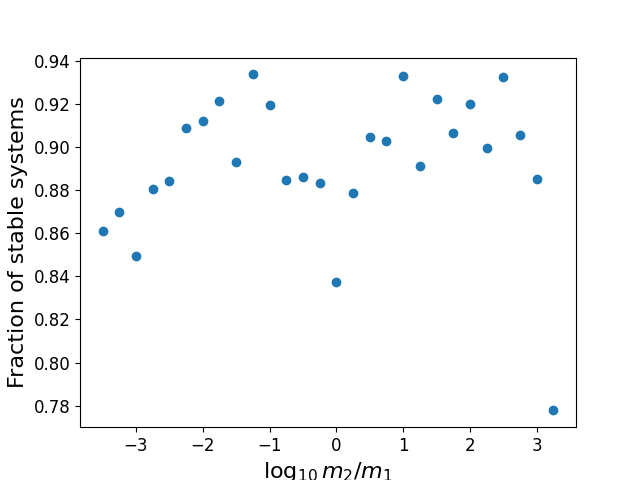}
    \caption{More results for the Nominal Kepler Analog Set. Top left: Scatter plot and histograms of all initial $\Delta$ and final $\Delta$ for stable systems (in blue) and unstable systems (in orange). Top right:
    Scatter plot and histograms of the planet multiplicity versus the initial $\Delta_{min}$ distribution stable systems (in blue) and unstable systems (in orange). The green triangles indicate the median of the initial $\Delta_min$ for stable systems.
    Bottom left: the fraction of stable systems versus planet multiplicity. Bottom right: the fraction of stable systems versus the mass ratio between the closest nearest-neighbor planet pair.}
    \label{fig:orig2}
\end{figure*}

\subsubsection{Initial vs final $\Delta$}

The final $\Delta$ values for the unstable planetary systems showed a wide dispersion, ranging from $\Delta < 0.001$ for neighboring planet pairs about to collide, to $\Delta \gtrsim 1000$ for an ejected planet in originally-neighboring planet pairs, as seen in the top left panel of Figure~\ref{fig:orig2}. For the long-term stable systems, the $\Delta$ distribution changes very little from the initial to the final state. This is an expected outcome since long-term stability is an indicator of few interactions and/or little orbital evolution that would change the planet pair spacings. In the unstable systems, a majority of the surviving planet pairs have final values of $\Delta$ near 10, similar to those of the stable systems.

\subsubsection{Planet multiplicity}

The stable systems with higher planet multiplicity had consistently smaller initial $\Delta_{min}$ values, as seen in the top right panel of Figure~\ref{fig:orig2} by the green triangles denoting the median of the $\Delta_{min}$ distributions for each planet multiplicity. The stable fraction also decreases monotonically with planet multiplicity, with a large fraction ($>80\%$) maintaining stability even at high multiplicities ($N\gtrsim8$), as seen in the bottom left panel of Figure~\ref{fig:orig2}.

\subsubsection{Planet pair mass ratio}

The bottom right panel of Figure~\ref{fig:orig2} shows a scatter plot of the fraction of stable systems versus the planet pair mass ratio. The stable fraction is found to be in the relatively narrow range of 0.78--0.94, with the highest stable fractions around $|\log_{10} m_2/m_1|\sim 1$. There is a weak trend of larger stable fraction in cases where the outer of the planet pair is more massive. There is also a hint of lower stable fraction in cases of nearly equal mass planet pairs and cases where the planet paris are very disparate in mass, $|\log_{10} m_2/m_1|\gtrsim 2$. There is large scatter in this plot so these trends are not conclusive.


\subsection{Very Closely Packed Set} \label{subsec:res_packed}

\begin{figure*}[ht]
    \centering
    {\Large \textbf{Stability in the Very Closely Packed Set}}\\
    \includegraphics[width=\columnwidth]{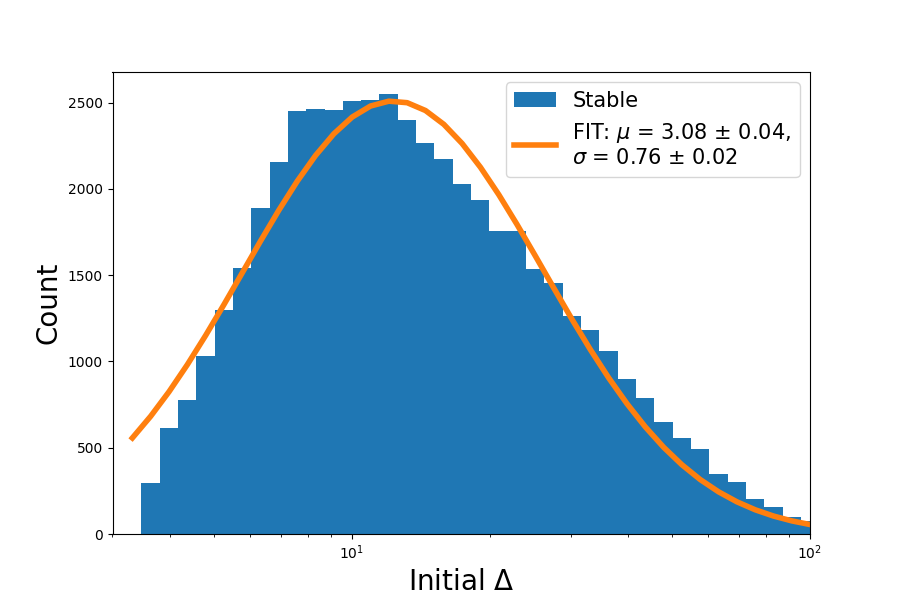}
    \includegraphics[width=\columnwidth]{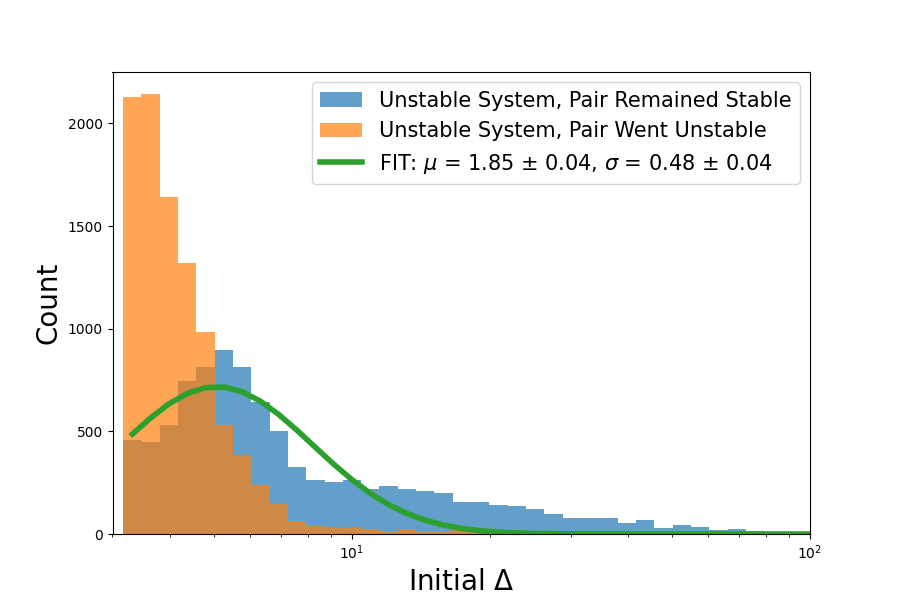}
    \includegraphics[width=\columnwidth]{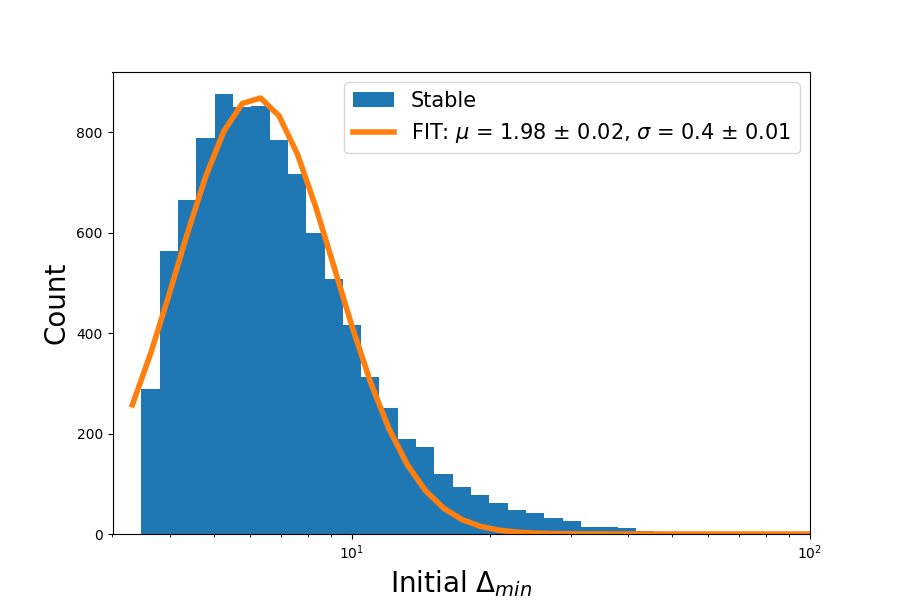}
    \includegraphics[width=\columnwidth]{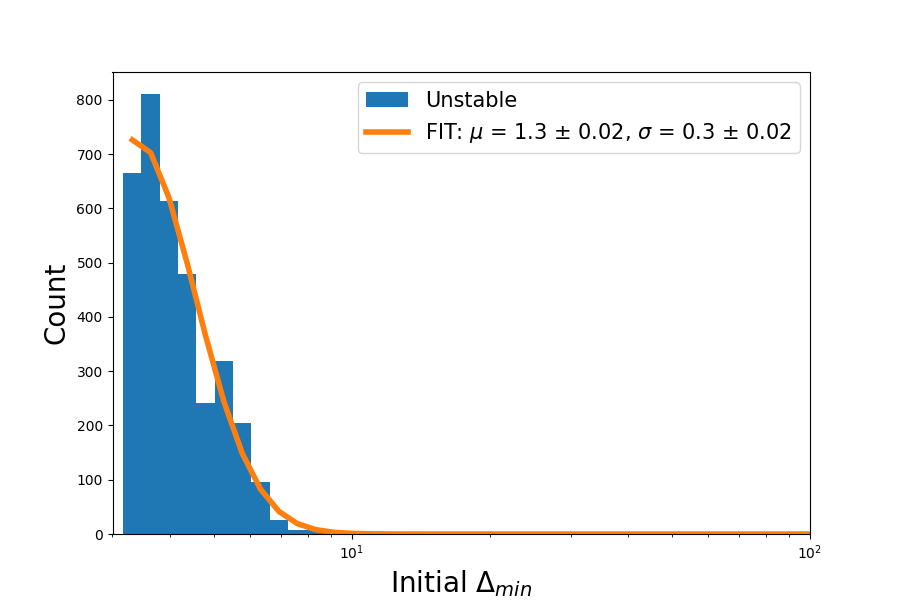}
    \caption{Results for the Very Closely Packed Set. Top: Histogram of all the $\Delta$ values for each nearest-neighbor planet pair in stable systems (left) and unstable systems (right). Bottom: Histogram of the initial $\Delta_{min}$ distribution and the best-fit log-normal function for stable systems (left) and unstable systems (right).}
    \label{fig:packed1}
\end{figure*}

For the Very Closely Packed Set (generated with a near-uniform distribution of $\log \Delta$ in the range $\log 3.5 - \log 10$), the initial $\Delta_{min}$ for stable systems also follows a roughly log-normal distribution, with parameters $\mu = 1.98\pm0.02$ and $\sigma = 0.40\pm0.01$. The initial $\Delta_{min}$ for unstable systems follow a log-normal distribution with parameters $\mu = 1.30\pm0.02$ and $\sigma = 0.30\pm0.02$ (see Figure~\ref{fig:packed1}). The stable distribution is very similar to that of the stable subset in the Nominal Kepler Analog Set, but the unstable distribution has a significantly smaller $\mu$ and larger $\sigma$. There are a few other notable differences that we discuss below.

\begin{figure*}[ht]
    \centering
    {\Large \textbf{Stability in the Very Closely Packed Set}}\\
    \includegraphics[width=\columnwidth]{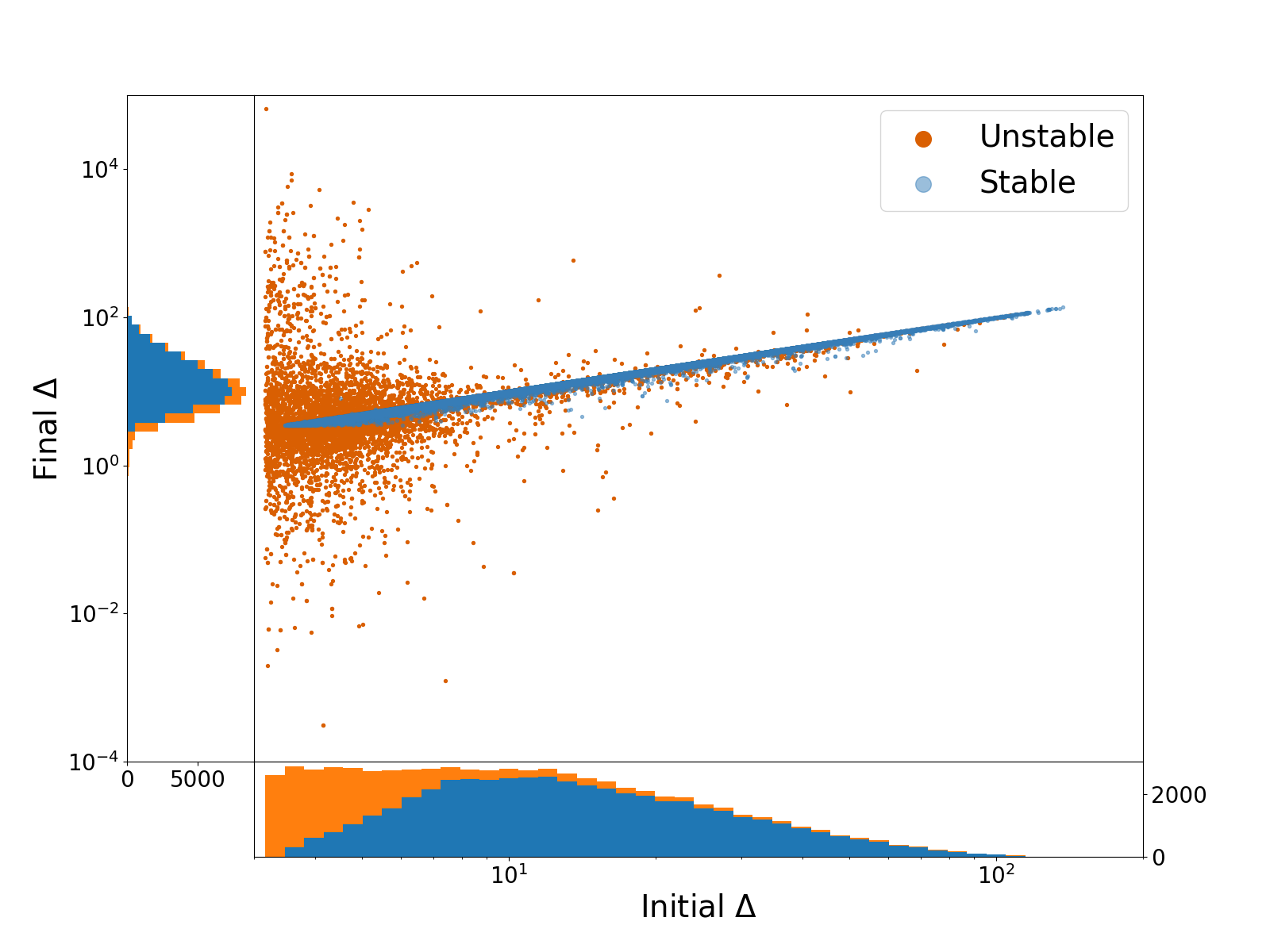}
    \includegraphics[width=\columnwidth]{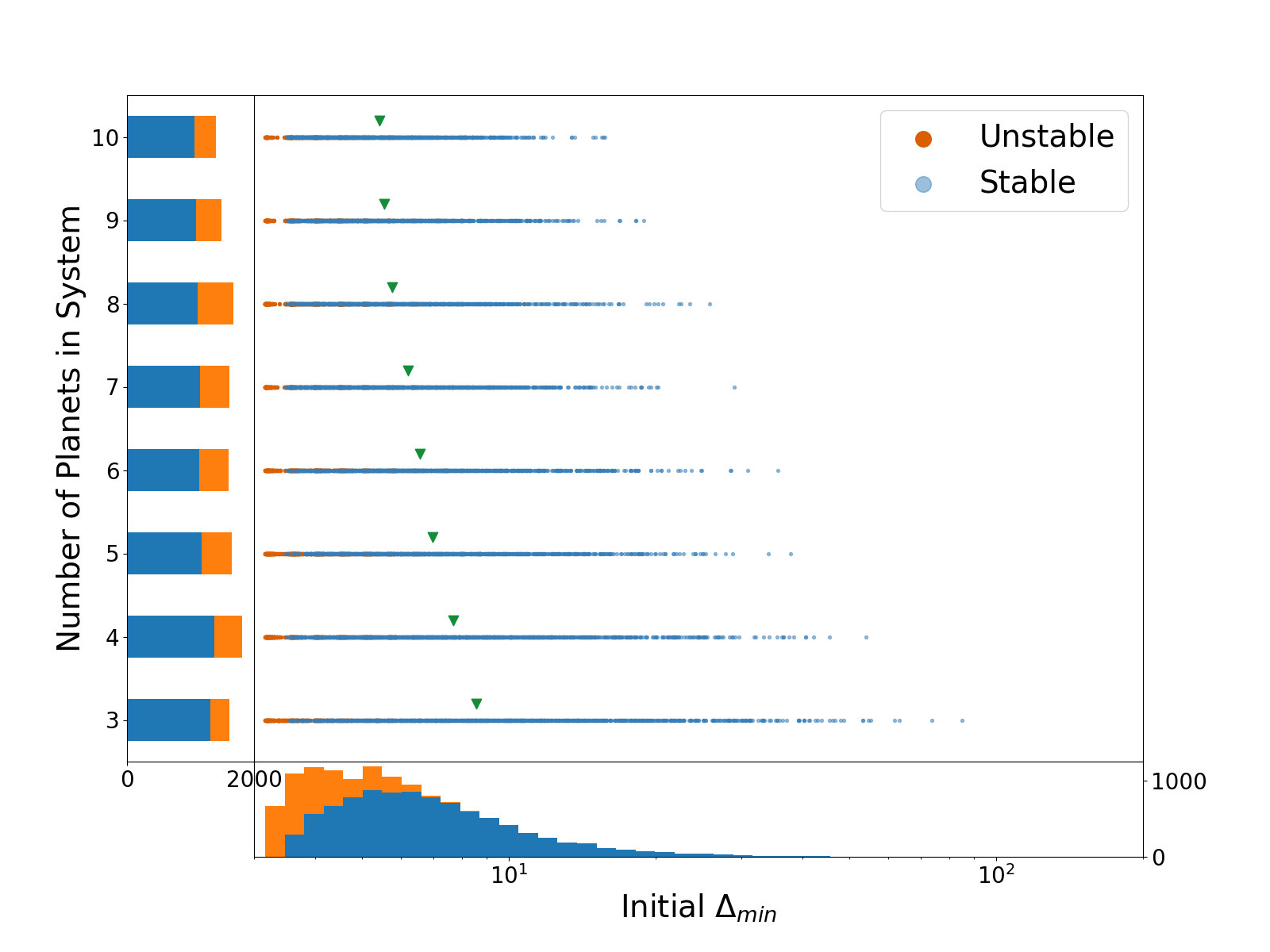}
    \includegraphics[width=\columnwidth]{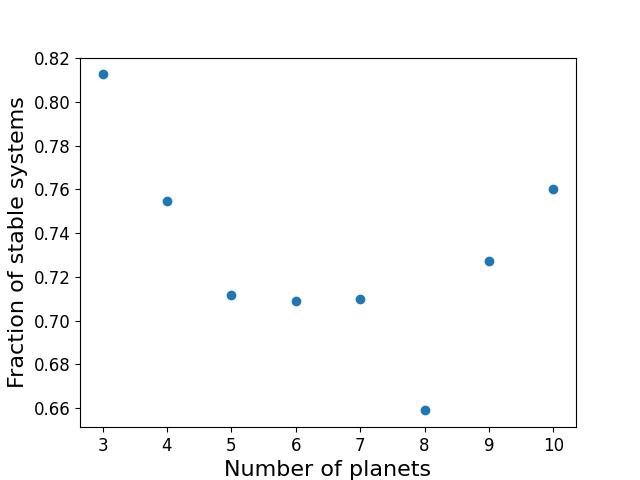}
    \includegraphics[width=\columnwidth]{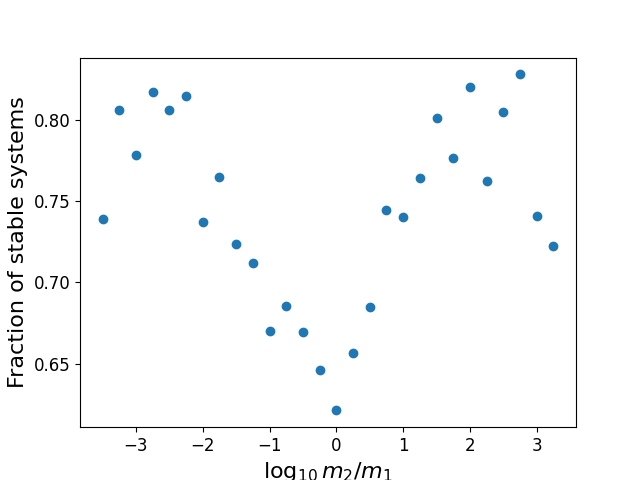}
    \caption{More results for the Very Closely Packed Set. Top left: Scatter plot and histograms of all initial $\Delta$ and final $\Delta$ for stable systems (in blue) and unstable systems (in orange). Top right:
    Scatter plot and histograms of the planet multiplicity versus the initial $\Delta_{min}$ distribution stable systems (in blue) and unstable systems (in orange). The green triangles indicate the median of the initial $\Delta_min$ for stable systems.
    Bottom left: the fraction of stable systems versus planet multiplicity. Bottom right: the fraction of stable systems versus the mass ratio between the closest nearest-neighbor planet pair. 
    }
    \label{fig:packed2}
\end{figure*}

\subsubsection{Initial vs final $\Delta$}

The comparison of all initial to all final $\Delta$ values shows the systems with a larger number of closely packed planets mostly tend to have planetary collisions (final values of $\Delta \ll 1$ for any neighboring planet pair) or ejections (final values of $\Delta \gg 10$ for any neighboring planet pair). There is an even wider dispersion of the final $\Delta$ values of the unstable systems in the Very Closely Packed Set, as can be seen in the top left panel of Figure~\ref{fig:packed2}, as the excess of close-separation planet pairs caused a higher number of encounters resulting in collisions or ejections.

\subsubsection{Planet multiplicity}

The trends in $\Delta_{min}$ with planet multiplicity have some similarities with those found in the Nominal Kepler Analog set, namely, lower initial $\Delta_{min}$ values for all planet pairs in systems with more planets (green triangles in the top right panel of Figure~\ref{fig:packed2}). However, in some contrast with the Nominal Kepler Analog set, the fraction of stable systems does not decrease monotonically in the Very Closely Packed Set (as seen in the bottom left panel of Figure~\ref{fig:packed2}); rather, it is flat for $N = 5-7$, followed by an increasing trend above $N=9$.

\subsubsection{Planet pair mass ratio}

In the scatter plot of the stable fraction versus planet pair mass ratio (bottom right panel of Figure~\ref{fig:packed2}, we observe several features in the Very Closely Packed Set that contrast with what we found for the Nominal Kepler Analog Set. The fraction of stable systems in the Very Closely Packed Set has a range of 0.60--0.85, which is lower and has little overlap with that found for the Nominal Kepler Analog Set; this is not surprising as a larger fraction of the closer packed planetary systems of the Very Closely Packed Set is unstable. Another interesting difference is that we observe a strong pattern in this scatter plot for the Very Closely Packed Set, in contrast with the significant scatter and only weak trends found in the Nominal Kepler Analog Set. The stable fraction is predominantly high for $|\log_{10} m_2/m_1| \approx 2-2.5$, prominently low in nearly equal mass planet pairs, and trends lower also in cases where the planet pairs are very disparate in mass, $|\log_{10} m_2/m_1| \gtrsim 2.5$. There is an evident nearly linearly increasing trend of the stable fraction as $|\log_{10} m_2/m_1|$ increases from $\sim 0$ to $\sim 2$.

\section{Discussion} \label{sec:discussion}

\subsection{The threshold of stability}

The primary goal of this analysis was to find a more accurate model for the  threshold of dynamical stability than a simple cutoff value of $\Delta_{min}$ somewhere greater than the minimum theoretical limit of $\Delta^*_{min} = 2 \sqrt{3}$ for the stability of a two planet system. In our procedure for generating initial states, we excluded systems with initial values of $\Delta_{min} < 2 \sqrt{3}$, as these would not have formed (or survived) in a stable region of the parameter space. We also classified as unstable any system which survived with $\Delta_{min} < 2 \sqrt{3}$ at the end of the integrations (which occurred in less than 0.1\% of all integrations), or if it did not pass the spectral fraction test for long term stability (which occurred in less than 5\% of all integrations). The resulting distributions of initial $\Delta_{min}$ are very similar between the Nominal Kepler Analog Set and the Very Closely Packed Set, and this similarity is true for both the stable systems and the unstable systems. The distributions of $\Delta_{min}$ for the two tested initial planetary spacing methods are log-normals of (1.97, 0.40) and (1.98, 0.40) for the stable systems and log-normals of (1.39, 0.13) and (1.30, 0.30) for the unstable systems. The unstable systems have a significant decrease in $\mu$ that defines the center of the distribution, as well as a significantly higher $\sigma$ denoting increased scatter.

We tested a factor of $\sim$$1.5$ more systems with closely separated planets in the second set, such that the initial distribution of $\Delta$ across all the systems was roughly uniform in log space from $\sim$$3.5-10$ (see Figure~\ref{fig:packed1}). Thus, the main difference in the number and fraction of stable vs. unstable systems comes from a higher number of these closer systems being found unstable. The nearly identical distributions of initial $\Delta$ values of the stable cases in the Nominal Kepler Analog Set and the Very Closely Packed Set provides evidence that the orbital spacings in the Kepler sample $\Delta_{Kepler}$ most likely arise from dynamical stability, rather than formation or migration, consistent with the results obtained by \citet[][]{Pu2015}.

The stable systems have a mean $\Delta_{min}$ of $\sim$$7.81$ and a median of $\sim$$7.03$, which is similar to but slightly smaller than the limiting value of $\Delta_{min} = 8$ adopted by the \texttt{SysSim} forward modelling framework \citep[][who also found $\Delta_{min} \sim 10$ in additional models]{He2019}. \citet[][]{Malhotra2015} noted a value of $\Delta_{min} \gtrsim 8$ for equal-mass planets in the low-planetary-mass regime ($\frac{m_p}{m_*} < 10^{-5}$) but $\Delta_{min} \approx 5$ for planets of higher mass ($\frac{m_p}{m_*} \sim 10^{-3}$), based on results from studies by \citet[][]{Chambers1996} and \citet[][]{Zhou2007}. The values reported by both \citet[][]{Malhotra2015} and \citet[][]{He2019} are consistent with our results across the entire set of mass ranges mentioned above, as our distribution of $\Delta_{min}$ peaks at $\sim$$5.69$. However, there is no evidence for a correlation between $\Delta_{min}$ and $m_p$ -- only an increase in instability -- when $\Delta_{min}$ in a system is between a neighboring pair containing a higher-mass planet.

Another value for $\Delta_{min}$ from the literature is $\Delta_{min} \approx 10-12$ for certain assumptions using the Kepler distributions \citep[][]{Pu2015}. We note that $\Delta = 10$ is at the 80th percentile of our $\Delta_{min}$ distribution and $\Delta = 12$ is at the 89th percentile; the result found by \citet[][]{Pu2015} is consistent with our distribution but significantly near the higher end. In particular, their study includes non-circular and non-coplanar starting orbits with a measured dispersion in eccentricity and mutual inclination, as well as a factor to measure stability over a set timescale. The dispersion in mutual inclinations and eccentricities induced an increase in the dynamical spacings measured \citep[Equation 14 of][]{Pu2015}, and the dynamical spacings for stability timescales at the length of our N-body integrations, we find a median stable spacing of $\sim$7 mutual Hill radii \citep[Equation 12 of][]{Pu2015}.

We find no significant difference (i.e., no significant evolution) between the initial and final $\Delta_{min}$ values for both of our sets of stable systems. We allowed encounters to freely change the orbital inclination and eccentricity of the planets, but since we did not use an initial seed of mutual inclinations or eccentric orbits, all of the planetary systems remained coplanar while a very large majority of the stable systems remained circular. Our results were not significantly different using a mutual inclination distribution ($\sim1-5^{\circ}$) and an eccentricity distribution ($\sim0.01-0.07$) that have been empirically found in known stable systems \citep[e.g.][]{He2020}. Therefore, we conclude there must be an additional factor that could describe why the $\Delta_{Kepler}$ distribution as measured by \citet[][]{Pu2015} has $\Delta_{min} \sim 10-12$ whereas our Nominal Kepler Analog Set has $\Delta_{min} \sim 6-7$. One such factor could be the similarity of planet masses within individual Kepler systems \citep[see e.g.,][]{Millholland2017, Weiss2018, Gilbert2020}, a correlation which is not well-preserved in either our Nominal Kepler Analog set or our Very Closely Packed Set.

\subsection{Multiplicity dependence}

We noticed in both sets of simulated planetary systems a modest trend of lower $\Delta_{min}$ with increasing planet multiplicity. In generating the initial suite of planetary systems as detailed in Section~\ref{subsec:init}, we did not place an upper limit on the outermost planetary period. Therefore, systems with more planets correspondingly had planets with longer orbital periods, and we did not place a restriction on the spacing that would artificially cause a higher number of low-separation pairs in systems with higher multiplicity. Thus, we are able to exclude as a possibility the potential observed degeneracy between high multiplicity and low planetary spacing, which would also have increased the strength of planetary perturbations and led to likely decreased stability \citep[see e.g.,][]{Pu2015, Rice2023}. Other studies have also shown a correlation between higher multiplicity and closer and more ordered spacings of planets \citep[see e.g.,][]{Wu2019}.

For stable planetary systems with 10 planets, the initial distribution of $\Delta_{min}$ had a median of $\sim$$5.5$ and a 95th-percentile value of $\sim$$12.5$, whereas for stable planetary systems with 3 planets the median was $\sim$$9.2$ and the 95th percentile was $\sim$$40$. This is shown in the middle-left panels of Figures~\ref{fig:orig2} (for the Nominal Kepler Analog Set) and~\ref{fig:packed2} (for the Very Closely Packed Set). The upper end of the range of initial $\Delta_{min}$ values for 3-planet stable systems almost reaches 100, while the upper end for 10-planet stable systems is less than 30. The medians of the initial $\Delta_{min}$ distribution for each planet multiplicity are denoted by the green triangles, which show a significant trend of decreasing initial $\Delta_{min}$ values with increasing planet multiplicity (note that the x-axis is in log scale for this figure).

Therefore, we infer the difference in $\Delta_{min}$ between low- and high multiplicity systems is due to the higher chance of having Jovian-mass planets in the system, creating lower $\Delta$ values. Since the planet masses were drawn log-uniformly from $10^{-6}$ to $10^{-3.5}$, higher multiplicity systems were more likely to contain one or more Jovian-mass planets. Any Jovian-mass planet in a system with similar period ratios will decrease the $\Delta$ value between it and any neighboring planets, so each Jovian-mass planet likely decreases two $\Delta$ values in a system. This makes the chance of every planet pair having a high $\Delta$ value much less likely for high planet multiplicity than for low planet multiplicity.

Even with the difference in $\Delta_{min}$ corresponding to multiplicity for the stable planetary systems, the upper bound of $\Delta_{min}$ for the unstable planetary systems in both subsets of data was relatively evenly spaced in multiplicity. The 95th-percentile value for the unstable cases in the Nominal Kepler Analog Set was $\sim$$5$, and this stayed relatively constant across all multiplicities. Similarly, the 95th-percentile value for the unstable systems in the Very Closely Packed Set was $\sim$$7$ for each planet multiplicity.

For the Nominal Kepler Analog Set, there was an increase in the fraction of unstable systems with increasing planet multiplicity. The unstable fraction for 3-planet systems was $\sim$4\% but for 9- and 10-planet systems was $\sim$15\%. This decrease in stability is consistent with the expectation that higher-multiplicity systems with close spacings are more likely to go unstable at an earlier time \citep[e.g.,][]{Pu2015, Petit2020}. However, for the Very Closely Packed Set, the rejection fraction seems at most weakly dependent on the number of planets in the system, as 3- and 10-planet systems had roughly the same unstable fractions that were the lowest amongst the different planet multiplicities.

\subsection{Mass ratio dependence}

We also wanted to determine if there were any correlations between the system stability and the mass ratios within the system. The random initialization of the systems with initial masses between sub-Earths and super-Jupiters allowed for a large absolute mass ratio between nearest-neighbor planets. Positive values of $\log \frac{m_2}{m_1}$ mean that the outer planet is more massive and negative values mean that the inner planet is more massive. This is intended to cover a wide variety of planetary systems. For both our samples of simulated planetary systems, we found a pattern between system stability and the mass ratios in the system. Systems with equal-mass planets are slightly less likely to be stable than those with unequal planetary mass ratios; however, as the mass ratios become more extreme (below $\sim\!0.01$ or above $\sim\!100$), the systems tend to become less stable (see the bottom right panels of Figures~\ref{fig:orig2}~and~\ref{fig:packed2}).

This pattern in stable fraction with planet mass ratio may have its origin in dynamical evolutionary pathways. \citet[][]{Goldberg2022} found that interactions in systems with planets of almost-equal mass tend to decrease the uniformity of the planets in the system by either removing planets from the system or in some cases merging together in a collision, increasing the mass inequality of initially similar-mass planets. In addition, \citet[][]{Lammers2023} showed that a level of intra-system mass uniformity, but specifically not rising to mass equality, is a likely result of dynamical interactions causing ejections and collisions during the planetary systems' evolution.

In general, the strength of the destabilizing effect scales proportionally with the mass of the perturbing bodies \citep[in this case the neighboring planet pair; see e.g.,][]{Volk2020}. For individual pairs of similar-mass planets, the destabilizing effect of a close interaction between the two planets is lessened if they have a slight but significant mutual inclination \citep[e.g.,][]{Rice2018}. In addition, unequal mass planets (especially with unequal spacings) tend to reach a larger part of the parameter space and to have a longer stability time than equal mass planets on equal spacings \citep[][]{Pu2015}.

\section{Summary} \label{sec:summary}

In this work, we carried out numerical simulations to study the distribution of the orbital separations in dynamically stable planetary systems. The critical parameter of interest is the dynamical separation (Eq.~\ref{e:Delta}), that is, the orbital separation in units of the mutual Hill radius of nearest-neighbor planets. We studied the overall distribution of $\Delta$ as well as that of its minimum value, $\Delta_{min}$, across nearest-neighbor pairs in multi-planet systems. We also studied how planet multiplicity as well as orbital period ratios and mass ratios affect the minimum value of this parameter. We generated two sets of simulated planetary systems: the Nominal Kepler Analog Set following the observed Kepler period distribution, and the Very Closely Packed Set containing more planet pairs with short separations. The key findings of our study are as follows:

\begin{itemize}
    \item The $\Delta_{min}$ distribution is not well fit by a single threshold value, a common assumption in previous studies. Instead, a log-normal function with parameters $\mu \approx 1.97\pm0.02$ and $\sigma \approx 0.40\pm0.02$ best fits the distribution of $\Delta_{min}$. This corresponds to a median value of $\Delta_{min}$ of $7.17\pm0.15$, a root-mean-square value $\langle\Delta_{min}^2\rangle^{\frac{1}{2}} = 8.4\pm0.3$, and the range $(4.8, 10.7)$ contains two-thirds of $\Delta_{min}$ values. The Nominal Kepler Analog Set and the Very Closely Spaced Set differ insignificantly from each other in the $\Delta_{min}$ distributions of their long-term stable systems.
    \item The overall distribution of $\Delta$ in long-term stable systems is well fit with a log-normal function with best-fit parameters $\mu=3.14\pm0.03$ and $\sigma=0.76\pm0.02$ (the Nominal Kepler Analog Set), and $\mu=3.08\pm0.04$ and $\sigma=0.76\pm0.02$ (the Very Closely Packed Set).
    \item The distribution of planetary spacings of the Kepler sample is consistent with sculpting by dynamical stability. Specifically, there is little difference between the overall $\Delta$ distributions found in the final stable architectures of planetary systems of the Nominal Kepler Analog Set and in the Very Closely Packed Set.
    \item There is a significant trend that long-term stable systems with high planet multiplicity have smaller initial minimum $\Delta$ values, which is attributed to the higher probability of a Jovian-mass planet in the system given our random planetary mass draws. 
    \item Planetary systems with large numbers of planets are in general less likely to be stable, but for the Very Closely Packed Set the long-term stable fraction is non-monotonic with the planet multiplicity, increasing again at the highest planet multiplicities ($N>8$).
    \item Systems with nearest-neighbor planet pairs having similar mass ratios tend to be less stable than those with somewhat dissimilar mass ratios, but the stable fraction tends to decrease again for mass ratios $\gtrsim 10^2$.
\end{itemize}

Our results on the statistical distribution function of the critical minimum value $\Delta_{min}$ of nearest-neighbor planet pairs in long-term stable planetary systems provide more precise and accurate parameters for planetary system evolution models, as well as improvements in statistically predictive tools, such as \textsc{Dynamite} \citep[][]{Dietrich2020, Dietrich2022}. 
Our results on the overall distribution of planetary spacings support the hypothesis that the $\Delta$ distribution of the thousands of exo-planetary systems discovered by the {\it Kepler} space telescope likely evolved from more tightly packed systems that underwent orbital evolution through dynamical instabilities (including ejections and collisions) to reach their current states. This suggests that dynamical interactions amongst planets decisively shape the architectures of compact planetary systems.

\begin{acknowledgments}
The results reported herein benefited from collaborations and/or information exchange within the program “Alien Earths” (supported by the National Aeronautics and Space Administration under Agreement No. 80NSSC21K0593) for NASA’s Nexus for Exoplanet System Science (NExSS) research coordination network sponsored by NASA’s Science Mission Directorate.  RM additionally acknowledges funding from NASA grant 80NSSC18K0397.  We acknowledge use of the software packages NumPy \citep[][]{Harris2020}, SciPy \citep[][]{Virtanen2020}, Matplotlib \citep[][]{Hunter2007}, and REBOUND \citep[][]{Rein2012, Rein2019}.  An allocation of computer time from the UA Research Computing High Performance Computing (HPC) at the University of Arizona is gratefully acknowledged.  The citations in this paper have made use of NASA’s Astrophysics Data System Bibliographic Services.
\end{acknowledgments}

\bibliography{main}{}



\end{document}